\documentclass[conference]{IEEEtran}
\IEEEoverridecommandlockouts
\usepackage{cite}
\usepackage{amsmath,amssymb,amsfonts}
\usepackage{algorithmic}
\usepackage{algorithm}
\usepackage{graphicx}
\usepackage{subcaption}
\usepackage{textcomp}
\usepackage{xcolor}
\usepackage{hyperref}
\def\BibTeX{{\rm B\kern-.05em{\sc i\kern-.025em b}\kern-.08em
    T\kern-.1667em\lower.7ex\hbox{E}\kern-.125emX}}
\begin{document}

\title{ffconnect: An Interactive Interface for Control Integration in Mid-Fidelity Wind Farm Simulation\\
% \thanks{This work is supported by the Natural Sciences and Engineering Research Council of Canada (NSERC).}
}

\author{\IEEEauthorblockN{Zekai Chen}
\IEEEauthorblockA{\textit{Department of Mechanical Engineering} \\
\textit{University of British Columbia}\\
Vancouver, Canada \\
zkaichen@student.ubc.ca}
\and
\IEEEauthorblockN{Casey Heiskell}
\IEEEauthorblockA{\textit{Department of Mechanical Engineering} \\
\textit{University of British Columbia}\\
Vancouver, Canada \\
caseyheiskell@gmail.com}
\and
\IEEEauthorblockN{Ryozo Nagamune}
\IEEEauthorblockA{\textit{Department of Mechanical Engineering} \\
\textit{University of British Columbia}\\
Vancouver, Canada \\
nagamune@mech.ubc.ca}}

\maketitle

\begin{abstract}
Wind farm control (WFC) plays a crucial role in mitigating the wake effect, the negative aerodynamic interactions among wind turbines. Recent advances in data-driven control and artificial intelligence offer new opportunities to design more intelligent WFC systems, motivating the need for a tool that supports interactive design and validation in simulation. To address this need, we present \texttt{ffconnect}, a general, open-source Python-based interface for \textit{FAST.Farm}, a mid-fidelity wind farm simulator. Compared to prior work, \texttt{ffconnect} introduces a restructured application programming interface (API) with enriched state access and supports integrating \textit{FAST.Farm} with modern scientific computing and machine learning ecosystems by building entirely on Python. In experiments, \texttt{ffconnect} shows negligible runtime overhead compared to the original \textit{FAST.Farm} across a range of simulation lengths and farm sizes, and demonstrates its effectiveness through a yaw-tracking case study. The source code of \texttt{ffconnect} can be found \href{https://github.com/ubccel/PyConnectFastFarm}{here}.
\end{abstract}

\begin{IEEEkeywords}
Wind Farm Control, Wake Effect, Mid-Fidelity Simulation, Message Passing Interface, Open-Source Software.
\end{IEEEkeywords}

\section{Introduction} \label{sec:intro}
Wind energy has become a cornerstone of the global energy transition owing to its ability to deliver large-scale power generation with limited carbon emissions~\cite{porte2020wind} and engineering maturity from large-scale grid integration~\cite{milligan2015alternatives}. Despite the huge potential, one major challenge that influences the competitiveness of wind energy is its high Levelized Cost of Energy (LCOE). Among all factors, one primary contributor to the high LCOE is the wake effect~\cite{WFFC_Review}. Wake effect refers to the negative inter-turbine aerodynamic interaction in wind farms. When turbines are placed in an array, upstream turbines extract kinetic energy from the incoming air, which creates regions of air with reduced wind speed and higher turbulence intensity behind them. Consequently, downstream turbines within these regions suffer from reduced power capture and increased structural loads~\cite{thomsen1999fatigue}. The wake effect impacts the performance of wind farms substantially. Annually, the wake effect may result in cumulative revenue losses ranging from $20\%$ to $30\%$~\cite{busby2012wind}. Therefore, mitigating the wake effect is a key priority for lowering the LCOE, which ultimately drives wind energy's competitiveness in the energy transition~\cite{kheirabadi2019quantitative}.

Wind Farm Control (WFC) is a key technology for mitigating the wake effect. Unlike conventional greedy control, where each turbine operates independently without accounting for its influence on neighboring turbines, WFC coordinates turbine operation globally to maximize whole-farm performance~\cite{WFFC_Review}. Previous studies that systematically review WFC methods have demonstrated that they substantially improve wind farm performance~\cite{kheirabadi2019quantitative, WFFC_Review}. Recent advances in data-driven control and artificial intelligence offer researchers new opportunities to develop more intelligent controllers, presenting significant potential when combined with existing WFC methods~\cite{dong2022wind, goccmen2025data}.

Since deploying on a real wind farm is prohibitively expensive, WFC development relies heavily on wind farm models for concept validation and fine-tuning~\cite{boersma2017tutorial}. An effective wind farm model captures two key aspects: (i) the aerodynamic interaction between the turbine rotor and the flow, and (ii) the evolution of the wake flow field across the wind farm~\cite{boersma2017tutorial}. Separated by the richness of wake dynamics they capture, wind farm models are broadly categorized into three groups: low, mid, and high-fidelity. Low-fidelity models such as \textit{FLOw Redirection In Steady-state (FLORIS)}~\cite{FLORIS} and \textit{FLOw Redirection and Induction Dynamics (FLORIDyn)}~\cite{FLORIDyn} are the most computationally efficient, allowing for online optimization. However, they typically resolve the wake dynamics only at steady-state, or with simplified quasi-dynamic features, limiting their applicability~\cite{boersma2017tutorial}. High-fidelity models such as \textit{Simulator fOr Wind Farm Application (SOWFA)}~\cite{SOWFA} capture the most complete wake dynamics by solving the three-dimensional Navier–Stokes (NS) equations using the large-eddy simulation~\cite{mehta2014large}. Nevertheless, they are the most computationally expensive options, making them impractical for controller design~\cite{boersma2017tutorial}. As a result, mid-fidelity models occupy the middle ground, offering a practical balance between computational speed and physical accuracy for control-oriented applications~\cite{boersma2017tutorial}.

% It is important to note that we do not intend to give a systematic introduction to wind farm wake models; a more comprehensive review can be found in G{\"o}{\c{c}}men et al.~\cite{goccmen2016wind} and Boersma et al.~\cite{boersma2017tutorial}. 

Among mid-fidelity options, \textit{Fatigue, Aerodynamics, Structures, and Turbulence Farm (FAST.Farm)}~\cite{fastfarm} is one of the most widely used tools due to its accuracy and efficiency. \textit{FAST.Farm} simulates each turbine using \textit{OpenFAST}~\cite{jonkman2024openfast} with turbine model \textit{FAST}~\cite{jonkman2005fast}, which provides richer turbine-level information than the actuator disk model employed by low-fidelity models such as \textit{FLORIS} and \textit{FLORIDyn}~\cite{moriarty2009wind}. At the farm level, wake evolution is resolved through the thin shear-layer approximation of the Reynolds-Averaged NS equations, and wake propagation is handled by the dynamic wake meandering model~\cite{larsen2007dynamic}, together capturing wake deficit, wake expansion, and wake deflection. Despite these strengths, \textit{FAST.Farm} presents practical limitations for advanced WFC design: WFC logic must be implemented in Fortran and recompiled for each change, which is cumbersome and poorly suited to an interactive development environment~\cite{smits2023fast}. Furthermore, Fortran lacks the machine learning and interactive development ecosystems available in Python, making it ill-suited for modern control algorithms such as Reinforcement Learning (RL) and game-theoretic methods~\cite{backus1978history}. Therefore, there is a need for a general, accessible Python-based interface for \textit{FAST.Farm} that supports the design and implementation of a wide range of WFC methods.

Several prior efforts have sought to address this gap. Smit et al.~\cite{smits2023fast} coupled \textit{FAST.Farm} to MATLAB/Simulink through the Message Passing Interface (MPI)~\cite{MPI}, enabling interactive design of WFC systems. However, both MATLAB and Simulink are commercial products requiring a paid license, which creates a problem for their accessibility. Moreover, the tool ships only skeleton controller templates that require substantial tuning before they produce stable closed-loop behavior. The Python-based variant of the interface provided by Monroc et al.~\cite{monroc2024wfcrl} was designed as part of a larger benchmark framework for WFC with RL. A particular focus was given to transfer learning between computationally cheap low-fidelity models (\textit{FLORIS}) and \textit{FAST.Farm}. However, the design was tightly bound to the RL given the design focus. Consequently, the information accessible to users is constrained, and adapting the framework to other control methods requires additional work to restructure the interface procedure.

To enhance the previous works, we present \textit{FAST.Farm} connector (\texttt{ffconnect}), an open-source Python interface to \textit{FAST.Farm}. We extend the aforementioned prior works in two regards:
\begin{itemize}
    \item \textbf{Optimized application programming interface (API) with enriched state access}: We restructure the software-in-the-loop logic into a cleaner, more flexible API. Moreover, we extend the set of available measurable signals from the handful of basic quantities accessible in~\cite{monroc2024wfcrl} to a rich state description. Signals such as structural loads, platform dynamics, and external atmospheric measurements may now be included in WFC logic.
    \item \textbf{Open-source Python implementation:} We release the code under an open-source license and allow users to leverage Python's extensive optimization, scientific computing, and machine-learning ecosystems to implement a wide variety of WFC methods.
    % \item \textbf{Foundation for data-driven control:} Our design directly supports the growing interest in learning-based control: by exposing the large volumes of data that high-fidelity simulations generate, \texttt{ffconnect} provides a foundation for developing more adaptive and intelligent wind farm controllers.
\end{itemize}

The structure of this paper is listed as follows: we introduce the structure of the \texttt{ffconnect} in Section~\ref{sec:structure}. In Section~\ref{sec:demo}, we validate the design by comparing the runtime overhead with the original \textit{FAST.Farm} and by demonstrating its effectiveness using a simple yaw-tracking case study. Finally, we conclude the paper in Section~\ref{sec:conclusion}.
\section{Interface Structure} \label{sec:structure}
This section introduces the fundamental structure of~\texttt{ffconnect} and outlines the general project workflow and simulation setup.

\subsection{Interface Structure}
Similar to prior studies, \texttt{ffconnect} relies on the Super Controller (SC)~\cite{fastfarm} and MPI to allow interactive WFC implementation in \textit{FAST.Farm}. The SC is a native \textit{FAST.Farm} component that encapsulates control logic and must be compiled into a dynamic link library (dll). In its standard form, the farm-level SC (\texttt{SC.dll}) receives measurements from individual turbines, computes control commands based on its embedded logic, and sends the computed commands to individual wind turbines by operating with turbine-level SC (\texttt{DISCON.dll})~\cite{smits2023fast}. A full description of the SC module is available in the official \textit{FAST.Farm} user guide~\cite{fastfarm}. In \texttt{ffconnect}, the \texttt{SC.dll} contains no WFC logic; it serves purely as a communication layer, collecting turbine measurements and forwarding calculated setpoints. Instead, all WFC logic is defined externally by the user, with MPI acting as the bridge between the external control logic and \textit{FAST.Farm}.

\begin{figure}[h!]
    \centering
    \includegraphics[width=\linewidth]{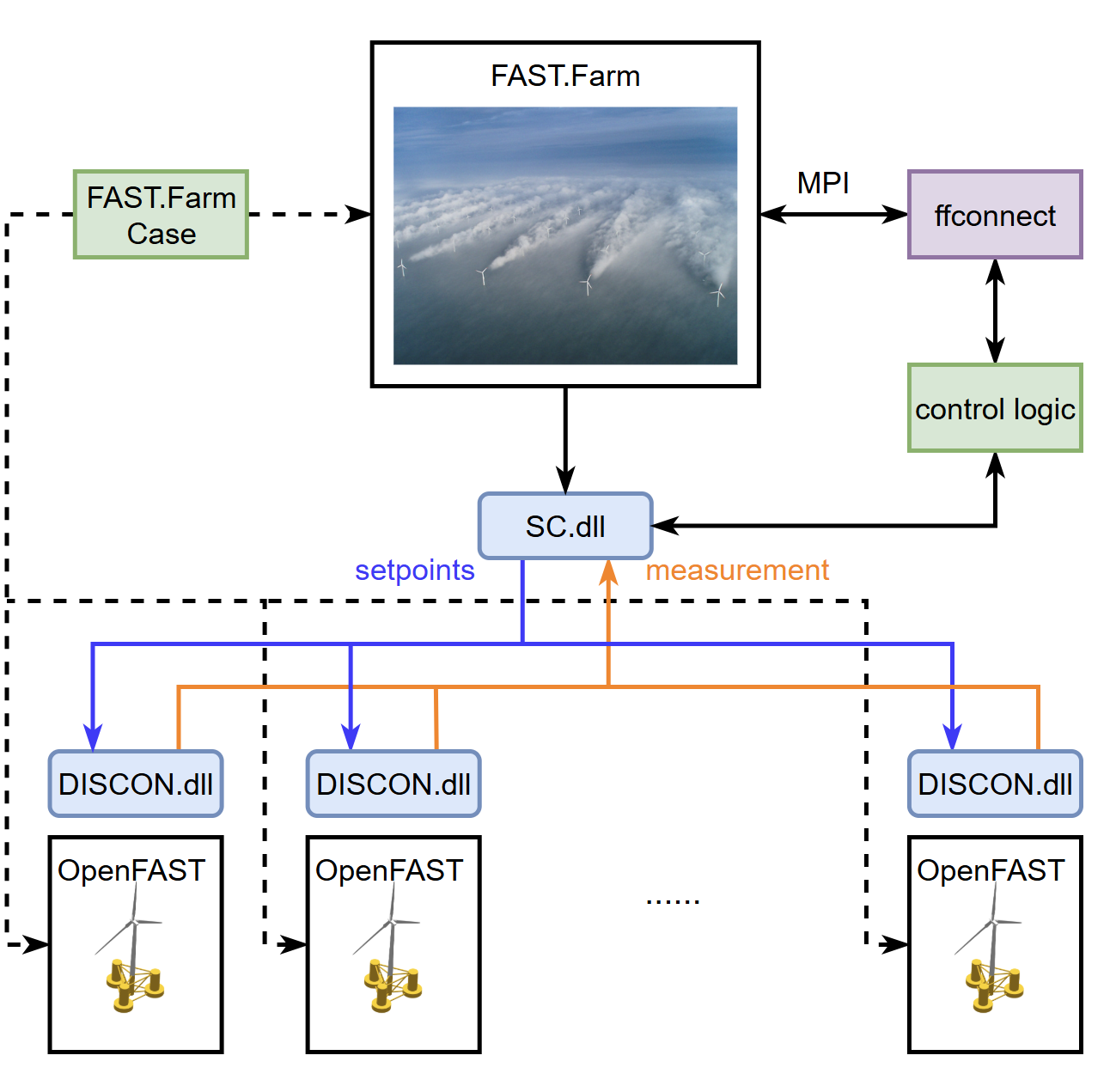}
    \caption{The structure of wind farm simulation integrated with~\texttt{ffconnect}. The wind farm picture is from Horns Rev offshore wind farm (Vattenfall) with visible wakes taken by photographer, Christian Steiness~\cite{hasager2013wind}.}
    \label{fig:structure}
\end{figure}

The overall structure of \texttt{ffconnect} is illustrated in Figure~\ref{fig:structure}, where colors distinguish different components: green represents user-defined components, orange represents signals from the wind turbines, and blue represents signals sent to the wind turbines. As the figure shown, the SC follows a hierarchical structure: \texttt{DISCON.dll} handles baseline turbine-level control~\cite{ROSCO} and operates the individual turbine to the required setpoints, while \texttt{SC.dll} manages communication between \texttt{DISCON.dll} and the \texttt{ffconnect} interface. 
% Notably, this hierarchy is designed to be modifiable so that \texttt{SC.dll} can be removed to accommodate future \textit{FAST.Farm} versions. 

When using \texttt{ffconnect}, users only need to define the \textit{FAST.Farm} simulation cases (.fstf files) using standard input files and specify the control logic in the simulation loop. The SC components are preconfigured and require no further setup by users. The \href{https://github.com/ubccel/PyConnectFastFarm}{Github repository} provides baseline controllers for both fixed-bottom and floating offshore wind turbines and supports nacelle-yaw ($\gamma$) and generator-torque ($\tau$) control, as well as collective and individual blade-pitch ($\beta$) control.

% Together, the SC hierarchy and MPI interface enable interactive, iterative simulation. Unlike the original \textit{FAST.Farm}, which requires all logic to be precompiled, \texttt{ffconnect} allows control logic — and other user-defined functionality — to be integrated and modified at runtime, offering substantially greater flexibility. The data workflow is described in detail in the following section.

\subsection{Interface Workflow}
\begin{figure}[h!]
    \centering
    \includegraphics[width=\linewidth]{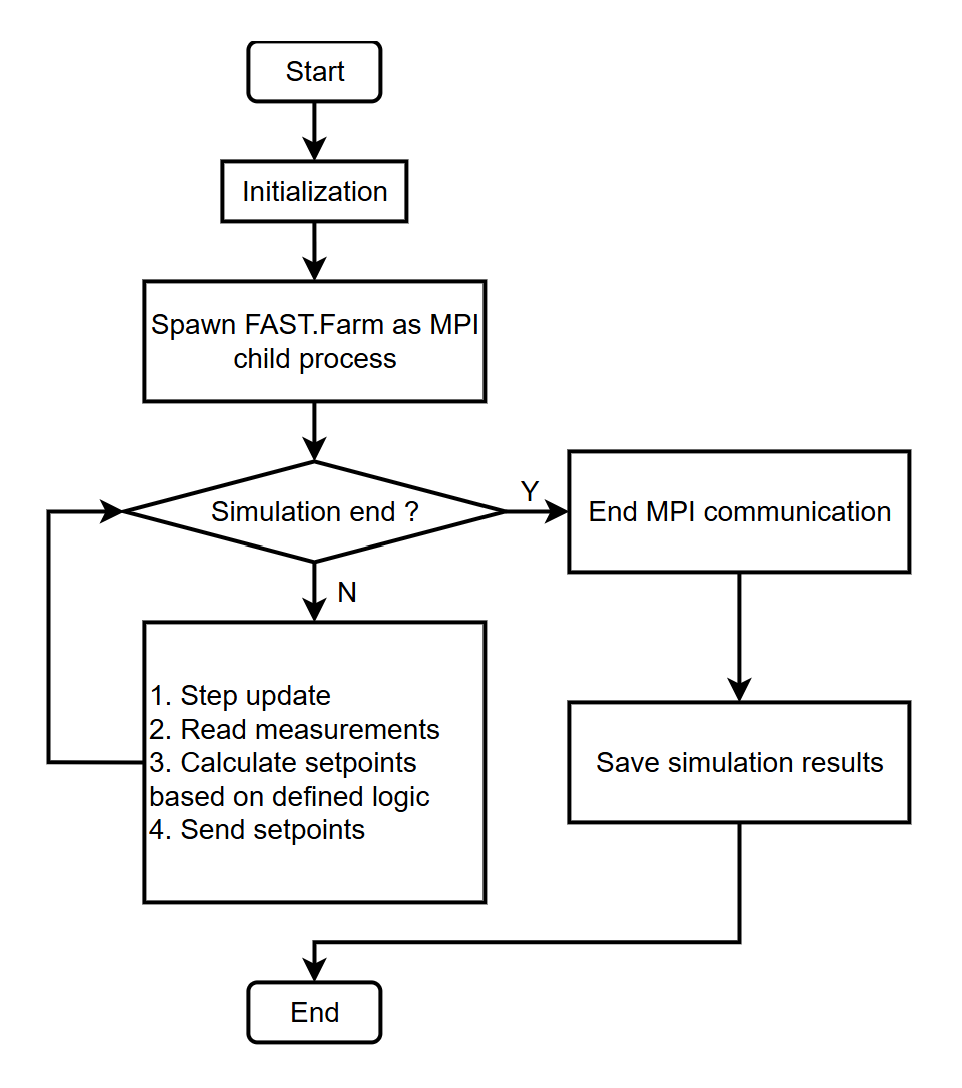}
    \caption{Flowchart that describes the workflow of \texttt{ffconnect}.}
    \label{fig:flowchart}
\end{figure}

Figure~\ref{fig:flowchart} illustrates the workflow of \texttt{ffconnect}. The simulation starts with initialization from the \textit{FAST.Farm} case file. Moreover, in order to enable interactive development, \texttt{ffconnect} spawns \textit{FAST.Farm} as an MPI child process. The simulation then proceeds through a user-defined loop consisting of four steps: 
\begin{enumerate}
    \item Advance the simulation by one timestep,
    \item Read the turbines' measurements,
    \item Calculate new setpoints according to the control logic,
    \item Send these setpoints back to the turbines.
\end{enumerate}
This loop repeats until the maximum number of iterations is reached, at which point the MPI communication is terminated, and the simulation ends. The user can store and post-process the results locally or access them through the standard \textit{FAST.Farm} and \textit{OpenFAST} output files.

The following pseudo-code of Algorithm~\ref{algo} shows the aforementioned procedure. In every iteration, the functions \texttt{get\_measure()} and \texttt{set\_command()} are responsible for retrieving measurements from and dispatching computed setpoints to individual wind turbines. To maximize flexibility, \texttt{ffconnect} allows the user to access measurements and control each wind turbine individually by specifying its index. To read a measurement, the user specifies the turbine index together with the name of the desired variable. To send setpoints, the user specifies the turbine index and provides the corresponding commands of $\gamma$, $\beta$, and $\tau$. The user may further choose which of these channels to actuate, while passing a value of \texttt{None} for a channel leaves it untouched, in which case the turbine's own baseline controller continues to govern that channel.

\begin{algorithm}
\caption{Pseudocode of \texttt{ffconnect} workflow}
\begin{algorithmic}[1]
\STATE Initialization
\WHILE{maximum iteration is not reached}
    \STATE \texttt{sim.step()} \algorithmiccomment{Update simulation step}
    \STATE \texttt{get\_measure(variable, turbine)}
    \STATE \algorithmiccomment{Implement control logic here to compute setpoints}
    \STATE \texttt{send\_command(yaw, pitch, torque, turbine)}
\ENDWHILE
\STATE Save data and stop
\end{algorithmic}
\label{algo}
\end{algorithm}
\section{Case Studies and Demonstrations} \label{sec:demo}
In this section, we evaluate \texttt{ffconnect} across two dimensions. First, we demonstrate that the MPI interface introduces negligible runtime overhead relative to the original \textit{FAST.Farm}, assessed over a range of simulation lengths and farm sizes. Second, we illustrate the end-to-end workflow through a yaw-tracking example.

\subsection{Runtime Overhead}
To assess whether the MPI functionality introduces significant runtime overhead, we compared the runtime of \texttt{ffconnect} and \textit{FAST.Farm}. Runtimes are evaluated across two dimensions: scaling with simulation length $T_\text{max}$ and scaling with wind farm size. In both cases, no WFC logic beyond the NREL baseline controllers~\cite{jonkman2024openfast} is active to keep the comparison fair. All the experiments are conducted on a laptop equipped with a 24-core Intel Ultra CPU and 32 GB of RAM, using the same version 3.5.5 \textit{FAST.Farm} executable throughout.

\begin{figure}[h!]
    \centering
    \includegraphics[width=\linewidth]{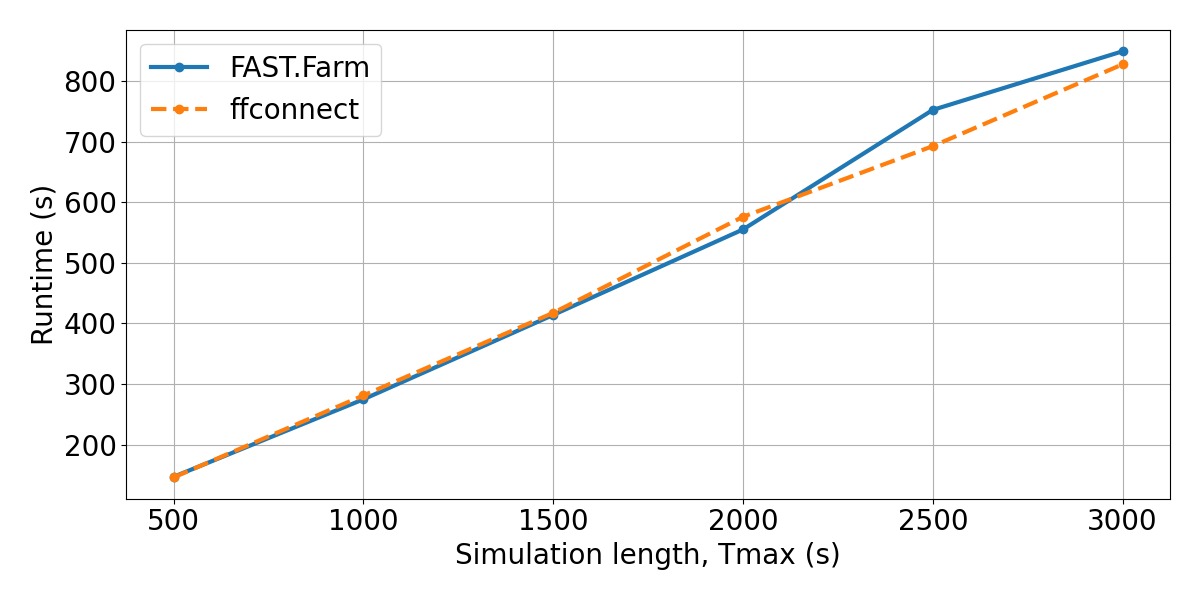}
    \caption{Comparison of runtime between \textit{FAST.Farm} and \texttt{ffconnect} with increasing $T_\text{max}$. The wind farm configuration is kept the same as $1\times2$ with $5D$ spacing for a fair comparison.}
    \label{fig:runtime_vs_simtime}
\end{figure}

Figure~\ref{fig:runtime_vs_simtime} presents the first comparison. A $1\times2$ array wind farm of two NREL~5MW turbines~\cite{NREL5MW} with 5 times rotor diameter $(5D)$ spacing is simulated at six cases with $T_{\text{max}}$ values in $\left\{500s, 1000s, \cdots, 3000s\right\}$. As shown, the runtimes of \textit{FAST.Farm} and \texttt{ffconnect} are nearly identical across all cases, indicating negligible overhead.

\begin{figure}[h!]
    \centering
    \includegraphics[width=\linewidth]{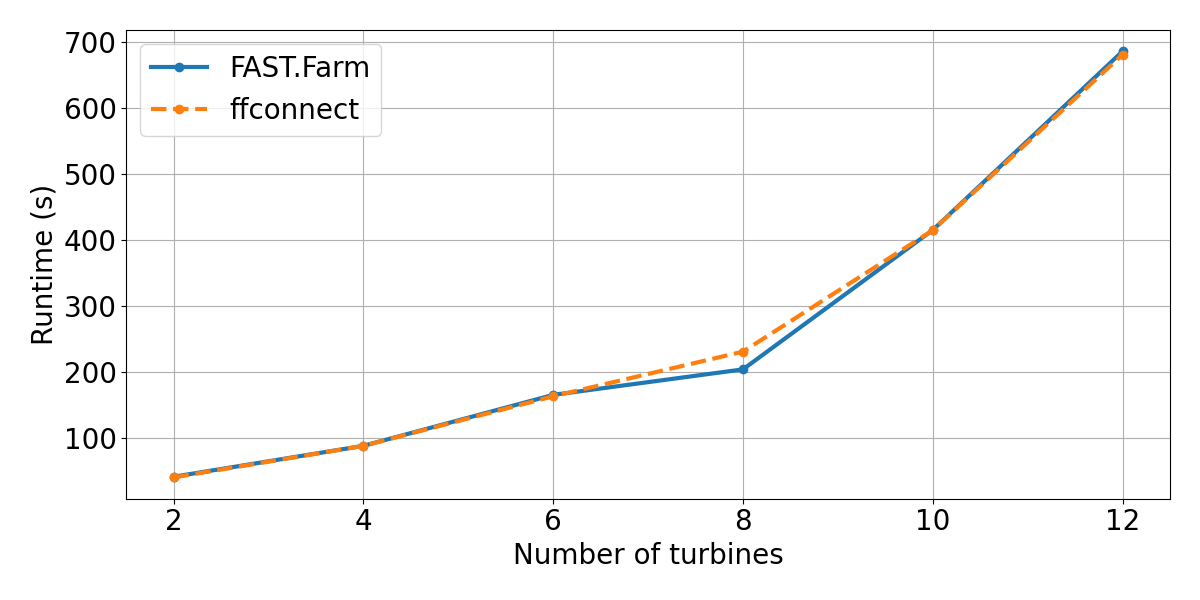}
    \caption{Comparison of runtime between \textit{FAST.Farm} and \texttt{ffconnect} with increasing wind farm size. The $T_\text{max}$ is kept the same as 300~seconds for a fair comparison.}
    \label{fig:runtime_vs_farmsize}
\end{figure}

Figure~\ref{fig:runtime_vs_farmsize} presents the second comparison, in which $T_\text{max}$ is fixed at $120s$ while farm size increases from $1\times2$ to $6\times2$. Again, the runtime difference between \textit{FAST.Farm} and \texttt{ffconnect} remain small across all configurations. Taken together, both experiments confirm that the MPI interface introduced by \texttt{ffconnect} adds negligible runtime overhead compared to the original \textit{FAST.Farm}. It is important to note that the inclusion of complex WFC logic will increase the runtime. Therefore, users are encouraged to consider code efficiency for implementations.

\subsection{Wind farm control case study}
In this section, we show the effectiveness of \texttt{ffconnect} through a yaw-tracking example, where each turbine tracks the wind direction via nacelle yaw control so that the rotor disk remains aligned with the incoming flow. While this example is intentionally simple, it exercises the complete \texttt{ffconnect} workflow end-to-end.

\begin{figure}[h!]
    \centering
    \includegraphics[width=\linewidth]{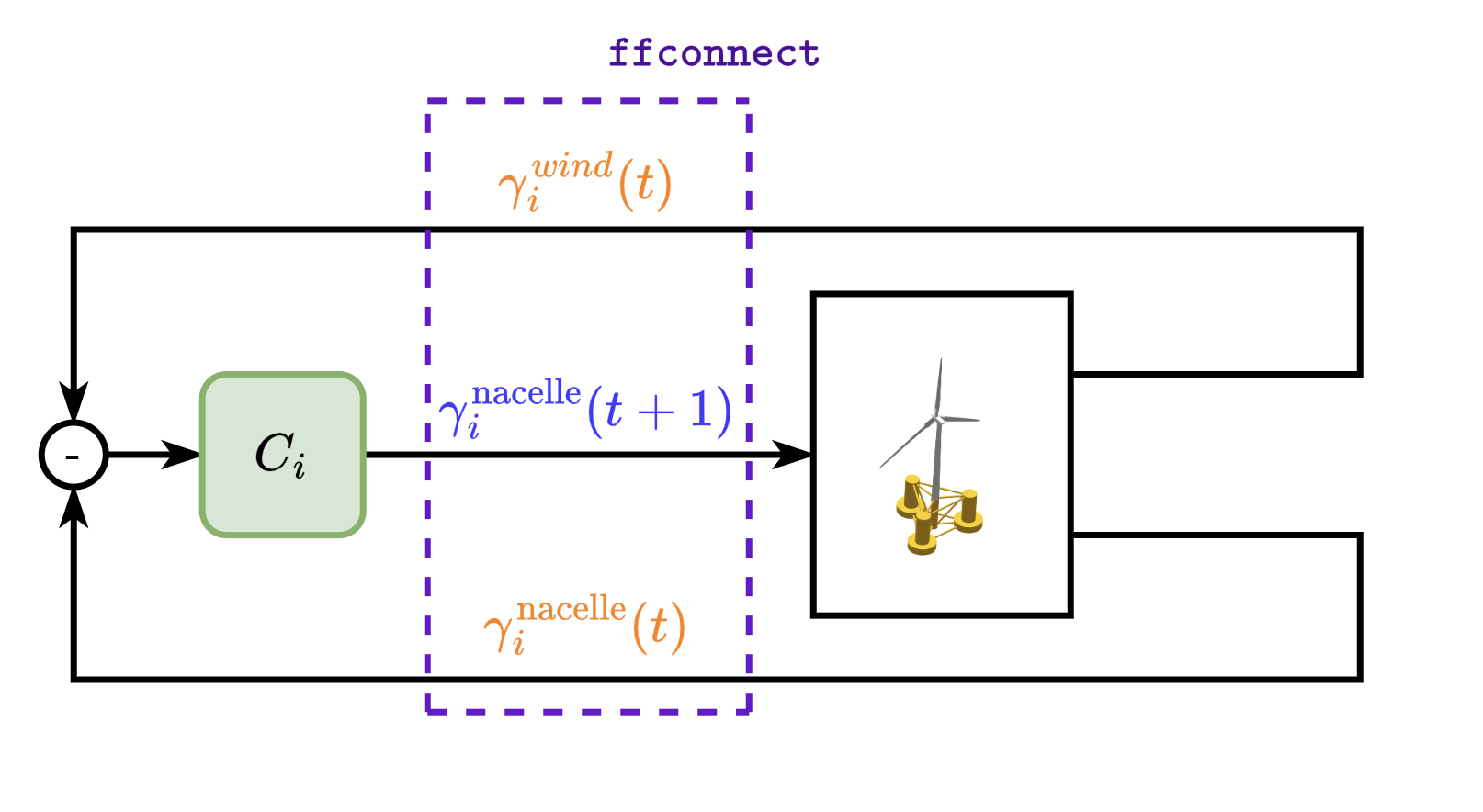}
    \caption{Block diagram of the nacelle yaw controller $C_i$ for each wind turbine. The color of different components is kept consistent with Figure~\ref{fig:structure}.}
    \label{fig:BD}
\end{figure}

The WFC logic requires each turbine to realign with the incoming wind as the wind direction changes. As illustrated in Figure~\ref{fig:BD}, at each iteration, the controller measures the wind direction $\gamma^\text{wind}_i(t)$ and the current nacelle yaw angle $\gamma_i^\text{nacelle}(t)$ for the $i^{th}$ wind turbine. Subsequently, the controller computes the yaw setpoints for the next timestep $t+1$ using a simple proportional controller with $K_p = 1$:
\begin{equation}
    \Delta\gamma^\text{nacelle}_i(t+1) = K_p \times (\gamma^\text{wind}_i(t) - \gamma_i^\text{nacelle}(t))
\end{equation}

\begin{figure}[h!]
    \centering
    \includegraphics[width=\linewidth]{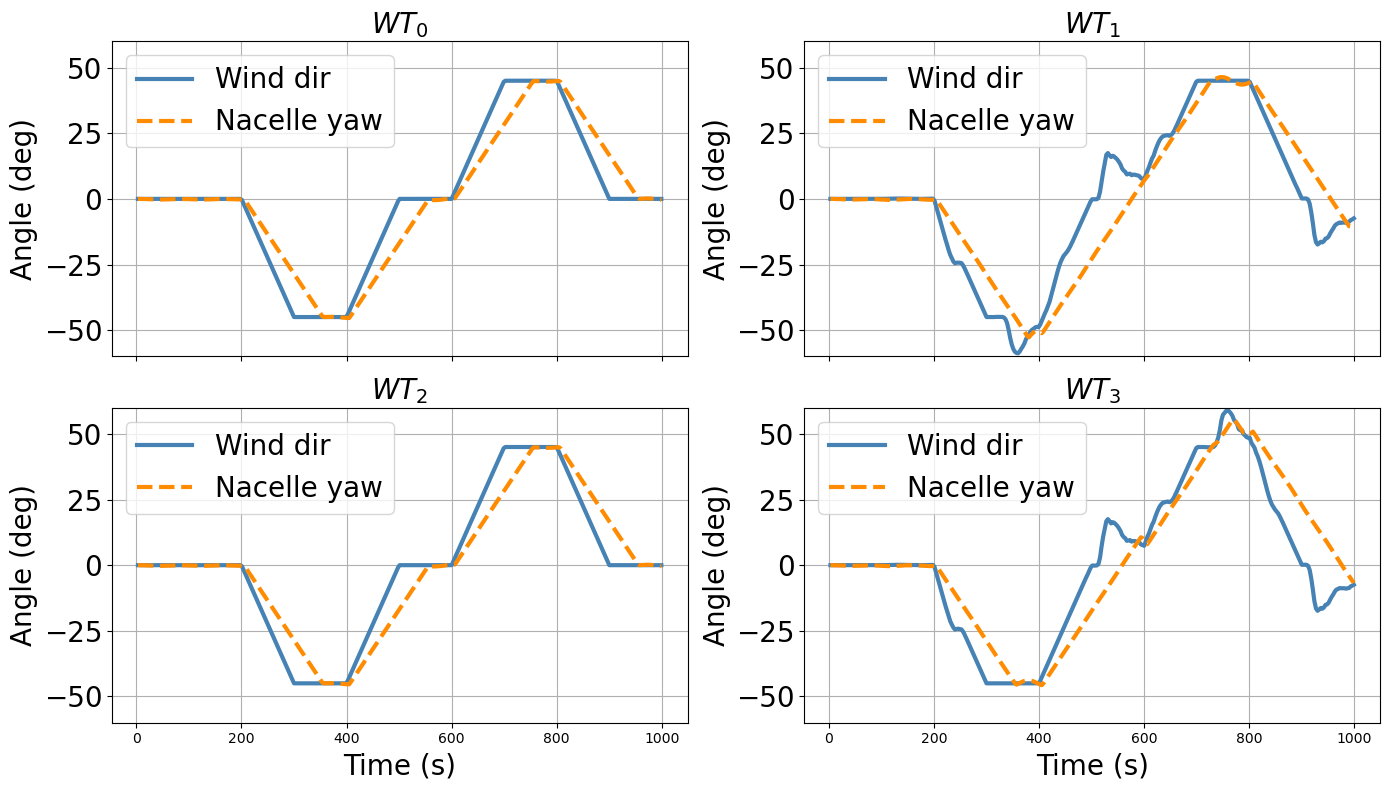}
    \caption{Results of the tracking performance of the $2\times2$ wind farm, showing good tracking results.}
    \label{fig:sim_result}
\end{figure}

The simulation is run on a $2\times2$ wind farm of four NREL 5~MW turbines~\cite{NREL5MW} with $5D$ spacing in both downwind and crosswind directions. The inflow speed is held constant at $10\:m/s$ at a reference height of $90\:m$, identical to the hub height of the NREL 5~MW turbine. Nevertheless, wind direction varies from $45^\circ$ to $-45^\circ$ in a ramp manner. The simulation is run for $1000\:s$ using \texttt{ffconnect}. Figure~\ref{fig:sim_result} shows the yaw-tracking result for each turbine. 

As the result shows, all turbines track the changing wind direction well, with the observed delay attributable to the $0.3\:\text{deg/s}$ yaw rate limit imposed by the NREL baseline controller for operational safety~\cite{NREL5MW}. From the figure, it can be observed that the wind direction measurements of the upstream wind turbines ($\text{WT}_0$ and $\text{WT}_2$) closely follow the defined wind profile, whereas those of the downstream turbine ($\text{WT}_1$ and $\text{WT}_3$) exhibit greater variability. This variability can be explained by the wake-induced flow disturbances from the upstream turbines in addition to the free-stream wind flow. 

\begin{figure*}
    \centering
    \begin{subfigure}[b]{0.32\textwidth}
        \centering
        \includegraphics[width=\linewidth]{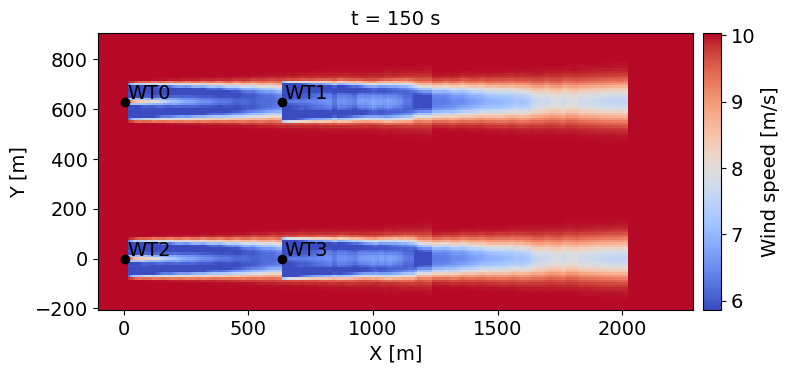}
        % \caption{First case}
        \label{fig:a}
    \end{subfigure}
    \hfill
    \begin{subfigure}[b]{0.32\textwidth}
        \centering
        \includegraphics[width=\linewidth]{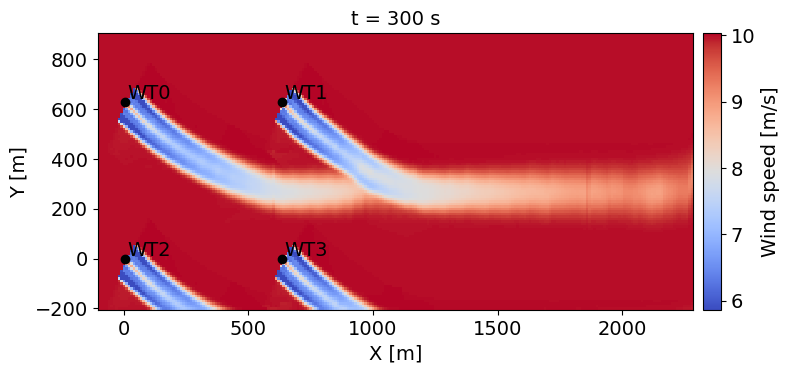}
        % \caption{Second case}
        \label{fig:b}
    \end{subfigure}
    \hfill
    \begin{subfigure}[b]{0.32\textwidth}
        \centering
        \includegraphics[width=\linewidth]{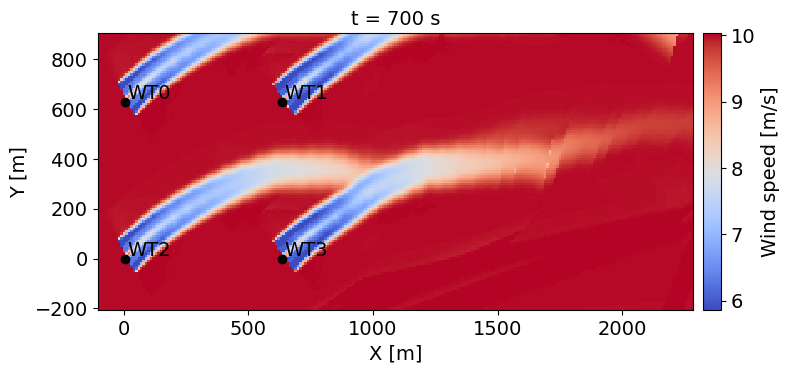}
        % \caption{Third case}
        \label{fig:c}
    \end{subfigure}
    \caption{Snapshots of the $2\times2$ wind farm in $\{150s, 300s, 700s\}$.}
    \label{fig:snapshots}
\end{figure*}

Finally, Figure~\ref{fig:snapshots} shows three simulation snapshots at $\{150s, 300s, 700s\}$, from which the dynamic movement of nacelle yaw can be seen.
% \begin{figure*}[h!]
%     \centering
%     \begin{subfigure}[b]{0.32\textwidth}
%         \centering
%         \includegraphics[width=\linewidth]{Figures/snapshot1.png}
%         % \caption{First case}
%         \label{fig:a}
%     \end{subfigure}
%     \hfill
%     \begin{subfigure}[b]{0.32\textwidth}
%         \centering
%         \includegraphics[width=\linewidth]{Figures/snapshot2.png}
%         % \caption{Second case}
%         \label{fig:b}
%     \end{subfigure}
%     \hfill
%     \begin{subfigure}[b]{0.32\textwidth}
%         \centering
%         \includegraphics[width=\linewidth]{Figures/snapshot4.png}
%         % \caption{Third case}
%         \label{fig:c}
%     \end{subfigure}
%     \caption{Snapshots of the $2\times2$ wind farm in $\{150s, 300s, 700s\}$.}
%     \label{fig:snapshots}
% \end{figure*}
\section{Conclusion} \label{sec:conclusion}

In this paper, we present an interactive simulation interface based on the message passing interface (MPI) and \textit{FAST.Farm} to allow wind farm control (WFC) system design and validation called \texttt{ffconnect}. To make the interface more accessible and general, we expanded previous works~\cite{smits2023fast, monroc2024wfcrl} by optimizing the application programming interface (API) logic with enriched state access. Runtime comparisons show that \texttt{ffconnect} only adds negligible runtime overhead compared to the original \textit{FAST.Farm}, and the following case study proves its effectiveness. While the presented example is intentionally simple, \texttt{ffconnect} is designed to support a wide range of WFC logic and seamless integration of \textit{FAST.Farm} with any external computational tool. Future work will focus on bypassing \texttt{SC.dll} to directly link \texttt{DISCON.dll} to the Python interface, enabling turbine-level control design.

\bibliographystyle{IEEEtran}
\bibliography{ref}

@article{porte2020wind,
  title={Wind-turbine and wind-farm flows: a review},
  author={Port{\'e}-Agel, Fernando and Bastankhah, Majid and Shamsoddin, Sina},
  journal={Boundary-layer meteorology},
  doi={10.1007/s10546-019-00473-0},
  volume={174},
  number={1},
  pages={1--59},
  year={2020},
  publisher={Springer}
}

@article{milligan2015alternatives,
  title={Alternatives no more: Wind and solar power are mainstays of a clean, reliable, affordable grid},
  author={Milligan, Michael and Frew, Bethany and Kirby, Brendan and Schuerger, Matt and Clark, Kara and Lew, Debbie and Denholm, Paul and Zavadil, Bob and O'Malley, Mark and Tsuchida, Bruce},
  journal={IEEE Power and Energy Magazine},
  doi={10.1109/MPE.2015.2462311},
  volume={13},
  number={6},
  pages={78--87},
  year={2015},
  publisher={IEEE}
}

@book{jonkman2005fast,
  title={FAST user's guide},
  author={Jonkman, Jason Mark and Buhl, Marshall L and others},
  volume={365},
  year={2005},
  publisher={National renewable energy Laboratory Golden, CO, USA}
}

@article{WFFC_Review,
	title = {Wind farm flow control: prospects and challenges},
	volume = {7},
	issn = {2366-7443},
	shorttitle = {Wind farm flow control},
	doi = {10.5194/wes-7-2271-2022},
	number = {6},
	journal = {Wind Energy Science},
	author = {Meyers, Johan and Bottasso, Carlo and Dykes, Katherine and Fleming, Paul and Gebraad, Pieter and Giebel, Gregor and Göçmen, Tuhfe and van Wingerden, Jan-Willem},
	month = nov,
	year = {2022},
	pages = {2271--2306},
}

@article{thomsen1999fatigue,
  title={Fatigue loads for wind turbines operating in wakes},
  author={Thomsen, Kenneth and S{\o}rensen, Poul},
  journal={Journal of wind engineering and industrial aerodynamics},
  doi={10.1016/S0167-6105(98)00194-9},
  volume={80},
  number={1-2},
  pages={121--136},
  year={1999},
  publisher={Elsevier}
}

@book{busby2012wind,
  title={Wind power: The industry grows up},
  author={Busby, Rebecca L},
  year={2012},
  publisher={PennWell Books}
}

@article{kheirabadi2019quantitative,
  title={A quantitative review of wind farm control with the objective of wind farm power maximization},
  author={Kheirabadi, Ali C and Nagamune, Ryozo},
  journal={Journal of Wind Engineering and Industrial Aerodynamics},
  doi={10.1016/j.jweia.2019.06.015},
  volume={192},
  pages={45--73},
  year={2019},
  publisher={Elsevier}
}

@article{dong2022wind,
  title={Wind farm control technologies: From classical control to reinforcement learning},
  author={Dong, Hongyang and Xie, Jingjie and Zhao, Xiaowei},
  journal={Progress in Energy},
  doi={10.1088/2516-1083/ac6cc1},
  volume={4},
  number={3},
  pages={032006},
  year={2022},
  publisher={IOP Publishing}
}

@article{goccmen2025data,
  title={Data-driven wind farm flow control and challenges towards field implementation: A review},
  author={G{\"o}{\c{c}}men, Tuhfe and Liew, Jaime and Kadoche, Elie and Dimitrov, Nikolay and Riva, Riccardo and Andersen, S{\o}ren Juhl and Lio, Alan WH and Quick, Julian and R{\'e}thor{\'e}, Pierre-Elouan and Dykes, Katherine},
  journal={Renewable and Sustainable Energy Reviews},
  doi={10.1016/j.rser.2025.115605},
  volume={216},
  pages={115605},
  year={2025},
  publisher={Elsevier}
}

@inproceedings{boersma2017tutorial,
  title={A tutorial on control-oriented modeling and control of wind farms},
  author={Boersma, Sjoerd and Doekemeijer, Bart M and Gebraad, Pieter MO and Fleming, Paul A and Annoni, Jennifer and Scholbrock, Andrew K and Frederik, Joeri Alexis and van Wingerden, Jan-Willem},
  booktitle={2017 American control conference (ACC)},
  doi={10.23919/ACC.2017.7962923},
  pages={1--18},
  year={2017},
  organization={IEEE}
}

@article{FLORIS,
  title={Wind plant power optimization through yaw control using a parametric model for wake effects—a CFD simulation study},
  author={Gebraad, Pieter MO and Teeuwisse, Floris W and Van Wingerden, JW and Fleming, Paul A and Ruben, Shalom D and Marden, Jason R and Pao, Lucy Y},
  journal={Wind Energy},
  doi={10.1002/we.1822},
  volume={19},
  number={1},
  pages={95--114},
  year={2016},
  publisher={Wiley Online Library}
}

@inproceedings{FLORIDyn,
  title={A control-oriented dynamic model for wakes in wind plants},
  author={Gebraad, Pieter MO and Van Wingerden, JW},
  booktitle={Journal of Physics: Conference Series},
  doi={10.1088/1742-6596/524/1/012186},
  volume={524},
  number={1},
  pages={012186},
  year={2014},
  organization={IOP Publishing}
}

@article{SOWFA,
  title={NWTC design codes-SOWFA},
  author={Churchfield, Matt and Lee, Sang},
  journal={URL: http://wind. nrel. gov/designcodes/simulator s/SOWFA},
  year={2012}
}

@book{fastfarm,
  title={Fast. farm user's guide and theory manual},
  author={Jonkman, Jason Mark and Shaler, Kelsey},
  year={2021},
  publisher={National Renewable Energy Laboratory Golden, CO, USA}
}

@article{jonkman2024openfast,
  title={OpenFAST/openfast: v3. 5.0},
  author={Jonkman, Bonnie and Mudafort, Rafael M and Platt, Andy and Branlard, E and Sprague, Mike and Ross, Hannah and Hall, Matt and Slaughter, Derek and Vijayakumar, Ganesh and Buhl, Marshall and others},
  journal={Zenodo},
  doi={10.5281/zenodo.7942867},
  year={2024}
}

@inproceedings{moriarty2009wind,
  title={Wind turbine modeling overview for control engineers},
  author={Moriarty, Patrick J and Butterfield, Sandy B},
  booktitle={2009 American Control Conference},
  doi={10.1109/ACC.2009.5160521},
  pages={2090--2095},
  year={2009},
  organization={IEEE}
}

@article{larsen2007dynamic,
  title={Dynamic wake meandering modeling},
  author={Larsen, Gunner C and Aagaard Madsen, H and Bing{\"o}l, Ferhat},
  year={2007}
}

@inproceedings{smits2023fast,
  title={A FAST. Farm and MATLAB/Simulink interface for wind farm control design},
  author={Smits, Coen-Jan and Silva, Jean Gonzalez and Chabaud, Valentin and Ferrari, Riccardo},
  booktitle={Journal of Physics: Conference Series},
  doi={10.1088/1742-6596/2626/1/012069},
  volume={2626},
  number={1},
  pages={012069},
  year={2023},
  organization={IOP Publishing}
}

@article{backus1978history,
  title={The history of Fortran I, II, and III},
  author={Backus, John},
  journal={ACM Sigplan Notices},
  doi={10.1145/960118.808380},
  volume={13},
  number={8},
  pages={165--180},
  year={1978},
  publisher={ACM New York, NY, USA}
}

@article{monroc2024wfcrl,
  title={Wfcrl: A multi-agent reinforcement learning benchmark for wind farm control},
  author={Monroc, Claire B and Bu{\v{s}}i{\'c}, Ana and Dubuc, Donatien and Zhu, Jiamin},
  journal={Advances in Neural Information Processing Systems},
  doi={10.52202/079017-4235},
  volume={37},
  pages={133254--133281},
  year={2024}
}

@article{hasager2013wind,
  title={Wind farm wake: The Horns Rev photo case},
  author={Hasager, Charlotte Bay and Rasmussen, Leif and Pe{\~n}a, Alfredo and Jensen, Leo E and R{\'e}thor{\'e}, Pierre-Elouan},
  journal={Energies},
  doi={10.3390/en6020696},
  volume={6},
  number={2},
  pages={696--716},
  year={2013},
  publisher={MDPI}
}

@article{ROSCO,
  title={A reference open-source controller for fixed and floating offshore wind turbines},
  author={Abbas, Nikhar J and Zalkind, Daniel S and Pao, Lucy and Wright, Alan},
  journal={Wind Energy Science},
  doi={10.5194/wes-7-53-2022},
  volume={7},
  number={1},
  pages={53--73},
  year={2022},
  publisher={Copernicus GmbH}
}

@techreport{NREL5MW,
  author       = {Jonkman, J and Butterfield, S and Musial, W and Scott, G},
  title        = {Definition of a 5-MW Reference Wind Turbine for Offshore System Development},
  institution  = {National Renewable Energy Laboratory (NREL), Golden, CO.},
  doi          = {10.2172/947422},
  url          = {https://www.osti.gov/biblio/947422},
  place        = {United States},
  year         = {2009},
  month        = {02}
}

@inproceedings{MPI,
  title={Message passing interface (mpi)},
  author={Barker, Brandon},
  booktitle={Workshop: high performance computing on stampede},
  volume={262},
  year={2015},
  organization={Cornell University Publisher Houston, TX, USA},
  URL={https://cac.cornell.edu/Education/training/StampedeJan2015/IntroMPI.pdf},
}

@article{mehta2014large,
  title={Large Eddy Simulation of wind farm aerodynamics: A review},
  author={Mehta, D and Van Zuijlen, AH and Koren, B and Holierhoek, JG and Bijl, H},
  journal={Journal of Wind Engineering and Industrial Aerodynamics},
  doi={10.1016/j.jweia.2014.07.002},
  volume={133},
  pages={1--17},
  year={2014},
  publisher={Elsevier}
}
\end{document}